\documentclass[12pt]{article}
\pdfoutput=1

\usepackage{a4}
\usepackage[fleqn]{amsmath}
\usepackage{hyperref}

\RequirePackage{setspace}
\RequirePackage{array}
\RequirePackage{booktabs}
\RequirePackage{mathrsfs}
\RequirePackage{amsmath,amssymb,amsxtra,amsthm}
\RequirePackage{braket}
\usepackage{cite}

\begin{document}

\begin{titlepage}

\begin{flushright}
ARC-18-20
\end{flushright}

\vskip 1cm
\begin{center}

{\Large{\textsc{GUT Scale Unification in Heterotic Strings
}}}

\vspace{1.5cm}
Carlo Angelantonj\,{}\footnote{\tt carlo.angelantonj@unito.it} and Ioannis Florakis\,{}\footnote{\tt iflorakis@uoi.gr}

\vspace{1cm}

{\it  ${}^1$ Department of Physics, University of Torino, \\
                    INFN Sezione di Torino and Arnold-Regge Centre\\
		Via Pietro Giuria 1, 10125 Torino, Italy \\ ~ \\
	 ${}^2$  Section of Theoretical Physics, Department of Physics,\\University of Ioannina, 45110 Ioannina, Greece 

}

\vspace{0.8cm}

\begin{abstract}
We present a class of heterotic compactifications where it is possible to lower the string unification scale down to the GUT scale, while preserving the validity of the perturbative analysis. We illustrate this approach with an explicit example of a four-dimensional chiral heterotic vacuum with $\mathcal{N}=1$ supersymmetry. 

\end{abstract}

\setcounter{footnote}{0}

\vspace{1.0cm}

\end{center}

\end{titlepage}

\pagestyle{empty}
\pagestyle{plain}

\pagenumbering{arabic}

\bibliographystyle{utphys}

The last few decades have been marked by striking progress both in formal developments of String Theory, as well as by efforts to classify string vacua that could serve as a basis for suitable phenomenological extensions of the Standard Model. In particular, the string effective action with $\mathcal N=1$ supersymmetry has been successfully reconstructed at tree level. Nevertheless, even the most promising string construction eventually requires a proper incorporation of quantum corrections before any quantitative contact with low energy phenomenology can be achieved.

The $F^2$ coupling in the effective action of the heterotic string is perhaps the most extensively studied such case in the literature, due to its direct relevance for the study of the running of gauge couplings in the framework of string unification, which has been one of the most elusive, yet desired, properties of String Theory. For the one-loop correction in the string coupling constant, and provided the string vacuum admits a description in terms of an exactly solvable worldsheet CFT, the calculation essentially amounts to computing the two-point function of gauge bosons on the Riemann surface of the worldsheet torus and separating out the logarithmic contribution of the massless states captured by field theory, from the heavy string states that are encoded into the threshold correction $\Delta_a$ \cite{Kaplunovsky:1987rp},
\begin{equation}
	\frac{16\pi^2}{g_a^2(\mu)} = k_a\frac{16\pi^2}{g_s^2} + b_a\,\log \left(\frac{\alpha}{4\pi^2}\frac{M_s^2}{\mu^2}\right) + \Delta_a \,.
	\label{runningCoupl}
\end{equation}
The running couplings $g_a(\mu)$ at scale $\mu$ associated to a gauge group factor ${\rm G}_a$, realised as a level-$k_a$ Kac-Moody algebra, are then related to the universal string coupling $g_s$ at tree level, while the massless states contribute via the familiar logarithmic term controlled by the corresponding beta function coefficient $b_a$.  Working in the $\overline{{\rm DR}}$ scheme results in the specific value for the constant $\alpha = 8\pi e^{1-\gamma}/3\sqrt{3}$, with $\gamma$ being the Euler-Mascheroni constant. Given the fact that the Planck mass does not renormalise in heterotic theories at any loop order \cite{Antoniadis:1992sa}, even in the case of spontaneously broken supersymmetry \cite{Florakis:2016aoi}, the string scale $M_s$ is conveniently related to the Planck scale by  $M_s=g_s M_\text{P}/\sqrt{32\pi}$.

The simplest properties of gauge thresholds are best visible in vacua that preserve $\mathcal N=2$ supersymmetry. The standard paradigm here is the compactification on ${\rm K3}\times T^2$ spaces, or their singular limits as $T^4/\mathbb Z_N\times T^2$ orbifolds, where $N=2,3,4,6$. With only a few assumptions requiring the factorisation of the $T^2$ and Kac-Moody lattices, essentially implying vanishing VEVs for the Wilson lines, the gauge thresholds become functions of the K\"ahler and complex structure moduli $T,U$ of the compactification 2-torus, respectively, and may be decomposed into the form \cite{Kiritsis:1996dn}
\begin{equation}
	\Delta_a^{\mathcal N=2} = -k_a \hat Y + b_a \hat\Delta \,.
	\label{N=2univ}
\end{equation}
The first term in this decomposition is a universal contribution due to the presence of the gravitational sector that depends only on the level $k_a$, but is otherwise independent of the charges of the states. The second term is instead proportional to the beta function coefficient of the $\mathcal N=2$ theory and encodes the contribution of the charged heavy states  running in the loop. Using arguments of modularity, holomorphy and the gravitational anomaly in six dimensions,  these contributions are completely determined\footnote{The expansion in the second line of the eq. \eqref{univY} is valid only in the large $T_2$ regime. For $T=U$ 
the whole expansion is outside the regime of validity and one should compute the integral using the methods developed in \cite{Angelantonj:2012gw} or numerically. In either case, the final expression is finite.} to be \cite{Kiritsis:1996dn}
\begin{equation}
\begin{split}
	\hat Y (T,U) &= \frac{1}{12}\int_{\mathcal F} \frac{d^2\tau}{\tau_2^2} \, \Gamma_{2,2}(T,U) \left( \frac{ \hat{\bar E}_2 \bar E_4 \bar E_6-\bar E_4^3}{\bar\eta^{24}}+1008\right) 
	\\
	&=\frac{1}{2} \, \log | j(T) - j (U) |^4 + \frac{4\pi}{3T_2} E (2;U) + O (e^{-2 \pi T_2})\,,
	\label{univY}
	\end{split}
\end{equation}
and \cite{Dixon:1990pc}
\begin{equation}
	\hat\Delta (T,U)= \int_{\mathcal F}\frac{d^2\tau}{\tau_2^2} \, \left( \Gamma_{2,2}(T,U) -\tau_2\right) = - \log \left[ \alpha \, T_2 U_2 |\eta(T)\eta(U)|^4\right] \,.
	\label{DKLint}
\end{equation}
The modular integrals are performed over the fundamental domain $\mathcal F = \{ \tau\in\mathbb C^+,~ |\tau|\ge1~ \rm{and}~ |\tau_1| \le 1/2 \}$ of the moduli space of complex structures $\tau=\tau_1+i\tau_2$ of the worldsheet torus. The modular invariant Narain lattice $\Gamma_{2,2}(T,U)$ of Lorentzian signature (2,2) encodes the contribution of Kaluza-Klein and winding excitations arising from the $T^2$ compactification, while $\eta(\tau)$ is the Dedekind $\eta$-function and $\hat{E}_2, E_4, E_6$ are the (quasi) holomorphic Eisenstein series of modular weight 2, 4 and 6, respectively. Furthermore $j (\tau )$ is the Klein invariant function and $E(s;z)$ is the real-analytic Eisenstein series. The subtraction of the linear term $\tau_2$ from the Narain lattice in \eqref{DKLint} simply amounts to removing the contribution of massless states which have  already been included into the logarithmic term of \eqref{runningCoupl}. 

It is remarkable that regardless of the specific details of the construction of the string vacuum, the gauge thresholds are decomposed according to the universal\footnote{Actually, there is a number of assumptions required for this simple universal form to hold. In general, this universality manifests itself in a more general form \cite{Mayr:1993mq,Angelantonj:2014dia,Angelantonj:2015nfa}.} $\mathcal N=2$ form  \eqref{N=2univ}.
This decomposition has very drastic physical consequences.  On the one hand, it leads naturally to the unification of all gauge couplings at the scale
\begin{equation}
	M_\text{U}= \frac{\sqrt{\alpha}}{2\pi}\,M_P\,g_s\exp(\hat\Delta/2)  \,,
\end{equation}
and  $g_s$ can be further expressed in terms of the common value  $g_\text{U}$ of the gauge couplings  at unification,
\begin{equation}
	g_s = g_\text{U} \left(1+\frac{g_\text{U}^2}{16\pi^2}\hat Y\right)^{-1/2}\,.
\end{equation}
On the other hand, if one adopts the {\it desert} doctrine and tries to set the string unification scale and coupling to their GUT values, $M_{\rm GUT}\sim 2\times 10^{16}\, {\rm GeV}$ and $g_{\rm GUT}^2 = 4\pi/25$, the precise expression of the gauge thresholds $\hat Y$ and $\hat \Delta$ reveals that no point in the $T,U$ moduli space is compatible with the unification data\footnote{See \cite{Dienes:1996du} and references therein for a review of previous works on GUT scale string unification.}. Indeed, the string unification scale in units of the GUT scale can be written as 
\begin{equation}
	\frac{M_\text{U}}{M_{\rm GUT}} =\frac{\sqrt{\alpha}}{4(2\pi)^{3/2}} \frac{ M_\text{P}}{M_{\rm GUT}}\, \frac{g_{\rm GUT}}{\sqrt{1+\frac{g_{\rm GUT}^2}{16\pi^2}\,\hat Y}}\, \exp(\hat \Delta/2)\,.
	\label{MUoverMGUT}
\end{equation}	
Since $M_\text{P}/M_{\rm GUT}\sim 6.1\times 10^2$, it is clear that the only way for the string unification scale to  match the GUT scale would be for the string thresholds $\hat Y$ and $\hat\Delta$ to take suitable values that exactly compensate this discrepancy. However, such a choice turns out to be impossible. In fact, the Narain lattice develops extrema \cite{Nair:1986zn} at points fixed under the T-duality  symmetry group ${\rm O}(2,2;\mathbb Z) = \text{SL} (2; \mathbb{Z})_T \times \text{SL} (2;\mathbb{Z})_U \ltimes \mathbb{Z}_2$, and  since the modular integrals \eqref{univY} and \eqref{DKLint} can be shown to be uniformly convergent over the full moduli space \cite{Angelantonj:2011br, Angelantonj:2012gw, Angelantonj:2013eja}, these points are also extrema for the thresholds $\hat Y$ and $\hat \Delta$. The minimum of the r.h.s. of  \eqref{MUoverMGUT} is achieved at the SU(3) point $T=U=e^{2\pi i/3}$, at which $\hat Y\sim 27.6$ and $\hat \Delta \sim 0.068$,  implying that the string unification scale still overshoots the GUT scale roughly by a factor of $20$.

A different issue arises when considering the case of large extra dimensions.  When $T_2={\rm Im}(T)$ becomes sufficiently large in string units, the associated Kaluza-Klein scale $M_{\rm KK}\sim 1/\sqrt{T_2}$ becomes much lower than the string scale or possibly even the GUT scale, while the string unification scale is pushed above the Planck scale exponentially fast. The reason is that in this limit, the thresholds asymptotically grow linearly with the volume of the compactification torus
\begin{equation}
	\hat\Delta \sim \frac{\pi}{3}T_2 \quad,\quad \hat Y \sim 4\pi T_2 \,. \label{decprob}
\end{equation}
Depending on the sign of the corresponding beta function coefficient $b_a$, the gauge coupling either decouples ($b_a>0$) or is driven to the non-perturbative regime ($b_a<0$). This is the so-called decompactification problem \cite{Antoniadis:1990ew} which reflects the fact that physics becomes effectively six dimensional above the KK scale and requires a linear dependence of the thresholds on the volume $T_2$. Technically, this linear growth in $T_2$ arises from the presence of 
\begin{equation}
\eta (\tau ) = q^{1/24} \prod_{n=1}^\infty (1-q^n )\,, \qquad j (\tau ) = \frac{1}{q} + 196884\, q + O(q^2 )\,,
\end{equation}
in the expressions for $\hat \Delta $ and $\hat Y$, with $q=e^{2\pi i\tau}$. Indeed, the combinations $T_2 |\eta (T )|^4$ and $j (T )$ are automorphic functions of the modular group $\text{SL} (2;\mathbb{Z} )_T$, reflecting the invariance of the thresholds under T-duality.

An obvious way to avoid the decompactification problem is to choose the moduli to take values sufficiently close to the string scale, although the mismatch between the string unification scale and the GUT scale still persists. 
Such a choice is always possible in supersymmetric vacua, where $T$ and $U$ are moduli and may be indeed treated as free parameters. However,  when supersymmetry is spontaneously broken, as for instance in the string realisation of the Scherk-Schwarz mechanism, the $T$ and $U$ scalars are no-longer moduli, since radiative corrections do generate a non-trivial potential. Under certain conditions \cite{Florakis:2016ani}, this potential  induces spontaneous decompactification of internal torii and drives the theory into the large volume regime where the decompactification problem becomes relevant.

At first sight, the mismatch between the string unification scale and the GUT scale on the one hand, and the decompactification problem at large volume on the other, appear to be entirely uncorrelated problems. The former relates to extrema of the Narain lattice  always occurring at values of the order of the string scale, while the latter is encountered as one approaches the boundary of moduli space. In this work, we propose that these two problems actually share a common origin, which may be traced back to the unbroken $\text{SL} (2; \mathbb{Z})_T \times \text{SL} (2;\mathbb{Z})_U \ltimes \mathbb{Z}_2$ T-duality symmetry of the Narain lattice.
Indeed,  asymptotically at large volume, $\Gamma_{2,2}(T,U)$ becomes independent of the integration variable $\tau$ and grows linearly with $T_2$, hence leading to the decompactification problem \eqref{decprob}.

Here, we wish to argue that  both problems may in fact be resolved simultaneously provided the T-duality group is broken in a such a way as to realise the replacement 
\begin{equation}
\text{SL} (2;\mathbb{Z} )_T \to \Gamma^1 (N) _T\,, \label{Tdbreaking}
\end{equation}
with $\Gamma^1 (N) _T$ being the congruence subgroup
\begin{equation}
\Gamma^1 (N) = \left\{ \begin{pmatrix} a & b \\ c & d \end{pmatrix}  \in \text{SL} (2;\mathbb{Z})  \, \Big|\, a,d = 1 \, (\text{mod} \, N), b= 0 \, (\text{mod} \, N)
\right\} \,,
\end{equation}
acting on the K\"ahler modulus.
Since the moduli dependence of the thresholds enters the amplitude only through  the KK and winding lattice sum, the breaking \eqref{Tdbreaking} implies that K3 and  $T^2$  no longer factorise,  but the $K3$ must be elliptically fibered over the $T^2$. The latter can be straightforwardly realised as an exactly solvable CFT in terms of freely-acting $\mathbb Z_N$ orbifolds, combining twists along the K3 directions together with shifts along a non-trivial cycle of the $T^2$. As a result, the amplitude for the factorised $\text{K3}\times T^2$ is replaced by
\begin{equation}
\int_\mathcal{F} \frac{d^2\tau}{\tau_2^2}\, \left( \frac{1}{N} \sum_{h,g \in \mathbb{Z}_N} \mathcal{A} \big[{\textstyle{h \atop g}}\big] \right)  \Gamma_{2,2} \to  \int_\mathcal{F} \frac{d^2\tau}{\tau_2^2}\, \frac{1}{N} \sum _{h,g \in \mathbb{Z}_N} \mathcal{A} \big[{\textstyle{h \atop g}}\big] \Gamma_{2,2} \big[{\textstyle{h \atop g}}\big] \,,
\end{equation}
where $h$ labels the orbifold sectors while the summation over $g$ implements the orbifold projection. Following \cite{Angelantonj:2013eja}, upon partial unfolding one can recast the r.h.s as
\begin{equation}
\frac{1}{N} \int_\mathcal{F} \frac{d^2\tau}{\tau_2^2}\, \mathcal{A} \big[{\textstyle{0 \atop 0}}\big] \Gamma_{2,2} \big[{\textstyle{0 \atop 0}}\big] +
\frac{1}{N} \int_{\mathcal{F}_N} \frac{d^2\tau}{\tau_2^2}\, \mathcal{A} \big[{\textstyle{0 \atop 1}}\big] \Gamma_{2,2} \big[{\textstyle{0 \atop 1}}\big] \,,
\end{equation}
where $\mathcal{F}_N$ is the fundamental domain of the Hecke congruence subgroup $\Gamma_0 (N)$. For the gauge thresholds, the helicity supertrace in $\mathcal{A} \big[{\textstyle{0 \atop 0}}\big]$ vanishes identically in accordance with the fact that the $(h,g) = (0,0)$ sector of the orbifold preserves all sixteen supercharges. Therefore the large volume behaviour of the thresholds depends on the way the coordinate shift is implemented in the Narain lattice. 
Although in string theory one has the freedom to shift a coordinate $X$ or its dual $\tilde X$, or even both, the requirement that the T-duality group contains $\Gamma^1 (N)_T$ automatically selects a geometric action, amounting to a momentum shift along a direction $\lambda_1 + \lambda_2 U$ of $T^2$, $\lambda_i \in \mathbb{Z}_N$. This results in the shifted Narain lattice
\begin{equation}
\Gamma_{2,2} \big[{\textstyle{0 \atop 1}}\big] (T,U) = T_2\, \sum_{m^i,n^i \in \mathbb{Z}} e^{-2 \pi i T \det (A)\,  - \frac{\pi T_2}{\tau_2 U_2} \left|  ( 1 \   U ) \, A \, \left( {\textstyle{\tau \atop 1}}\right)\right|^2  }  \,, \label{Narainshifted}
\end{equation}
with 
\begin{equation}
A = \begin{pmatrix} n^1 & m^1 + \frac{\lambda_1}{N} \\ n^2 & m^2 + \frac{\lambda_2}{N} \end{pmatrix} \,.
\end{equation}
For this construction, the $\hat \Delta$ and $\hat Y$ contributions to the thresholds take the generic form
\begin{equation}
\hat \Delta = \int_{\mathcal{F}_N} \frac{d^2 \tau}{\tau_2^2}\, \Gamma_{2,2} \big[{\textstyle{0 \atop 1}}\big] (T,U)\,,
\qquad
\hat Y = \int_{\mathcal{F}_N} \frac{d^2 \tau}{\tau_2^2}\, \Gamma_{2,2} \big[{\textstyle{0 \atop 1}}\big] (T,U)\, \Phi_N (\tau )\,,
\end{equation}
with $\Phi_N$ being a weak quasi-holomorphic modular form of $\Gamma_0 (N)$ with vanishing constant term. It can be shown  \cite{Angelantonj:2013eja, Angelantonj:2015rxa} that the dominant behaviour of the thresholds at large-$T_2$  is
\begin{equation}
	\hat\Delta \sim -\log \left( \alpha f_N (U)\, T_2 \right) + O(e^{-2\pi T_2})  \quad,\quad \hat Y \sim  O(T_2^{-1}) \,,
	\label{asymptBehav}
\end{equation}
where $f_N (U)$ is a function of the complex structure modulus $U$, whose explicit expression depends on the order $N$ of the freely-acting orbifold. Contrary to the case where the T-duality group is maximal, the replacement \eqref{Tdbreaking} forces the threshold $\hat \Delta$ to become unbounded from below. Therefore, independently of the new extrema of $\hat \Delta$, one may always find suitable values of the volume $T_2$ to match the string unification and GUT scales. In fact, imposing $M_\text{U} = M_\text{GUT}$ in  \eqref{MUoverMGUT} one finds
\begin{equation}
T_2 \simeq \frac{ g_\text{GUT}^2}{128\pi^3 f_N (U)}\, \left( \frac{M_\text{P}}{M_\text{GUT}}\right)^2 \sim 50\,,
\end{equation}
where,  we have assumed that $f_N (U) = O(1)$ for string-scale values of the $U$ modulus, which is typically the case for orbifold constructions, where the discrete symmetry is only compatible with fixed values of $U$.

Although the idea of using freely-acting orbifolds which restore $\mathcal{N}=4$ supersymmetry at large volume as a method to eliminate the decompactification problem has already been discussed in \cite{Kiritsis:1996xd, Faraggi:2014eoa}, the fact that this same technique simultaneously allows one to lower the string unification scale arbitrarily close to the GUT scale  had not been previously considered. 

The paradigm we have exposed so far actually enjoys unbroken $\mathcal{N}=2$ supersymmetry in four-dimensions, and is thus incompatible with the chiral nature of the observed spectrum of elementary particles.  For this reason, any attempt to apply our program to string vacua with sensible  low-energy phenomenology requires an extension of the above analysis to chiral $\mathcal{N}=1$ orbifold compactifications. In such vacua,  threshold corrections decompose as
\begin{equation}
	\Delta_a = d_a + \sum_i \left(\beta_{a,i}\,\hat\Delta^{(i)} -k_a \hat Y^{(i)}\right)\,, \label{Noneth}
\end{equation}
where $d_a$ are moduli independent constants originating from $\mathcal N=1$ subsectors, while $\hat\Delta^{(i)}$ and $\hat Y^{(i)}$ are moduli-dependent contributions of the $i$-th $\mathcal N=2$ subsector. The associated $\mathcal N=2$ beta function coefficients are denoted $\beta_{a,i}$ and may be related to six-dimensional anomaly coefficients \cite{Derendinger:1991hq}.

Achieving gauge coupling unification (at any scale) is no longer automatic but requires some  additional conditions on the charged spectrum. Namely, the quantity 
\begin{equation}
	\Phi_a\equiv \frac{b_a}{k_a}\log\left(\frac{\alpha}{4\pi^2}\,\frac{M_s^2}{M_U^2}\right)+\frac{d_a}{k_a}+\sum_{i} \frac{\beta_{a,i}}{k_a}\,\hat\Delta^{(i)} \,,
	\label{PhiCond}
\end{equation}
must be the same for all unified gauge group factors, \emph{i.e.} $\Phi_a=\Phi_b$ for all $a,b$, and reduces to the conditions discussed in \cite{Ibanez:1992hc} in the special case $d_a=0$ and $\Phi_a=0$. 
 It should be stressed that these conditions are especially relevant for mirage-like unification, where the gauge group in the full string construction is already broken down to the Standard Model group factors. In the case of  `true' unification, instead, there is only one relevant gauge group above the unification scale and, therefore, the previous conditions become redundant and it is easy to choose the moduli $T_i$ to match the GUT data.   Nevertheless, our analysis is quite general and may be applied in both cases\footnote{In general, it may be necessary to further generalise \eqref{Tdbreaking} to include also the group $\Gamma_1(N)_T$ for certain $\mathcal N=2$ subsectors in order to allow values $T_{i,2}<1$, which may be easily obtained by replacing the corresponding momentum shifts  by winding shifts.}. In the mirage scenario with at most three gauge group factors, the conditions $\Phi_a=\Phi_b$ with $M_\text{U}=M_{\rm GUT}$ can always be satisfied for a suitable choice of the moduli $T_i$.

It is a well-known fact that heterotic vacua with $\mathcal{N}=1$ supersymmetry can be realised as orbifold limits $T^6 / \Gamma$ of Calabi-Yau spaces, with $\Gamma$ a  suitable discrete symmetry group of the compactification and charge lattices, preserving exactly four Killing spinors. In such vacua, thresholds are typically moduli-independent, unless  $\Gamma$ contains elements which preserve eight supercharges, corresponding to sectors  enjoying enhanced $\mathcal{N}=2$ supersymmetry. 
Thresholds then decompose according to eq. \eqref{Noneth}, and the decompactification problem becomes relevant. Therefore, the  breaking pattern \eqref{Tdbreaking} ought to be implemented for all such sectors. Special care is required so that chirality be not spoiled by this procedure.  

In fact,  in the simplest instances of $\mathbb{Z}_2 \times \mathbb{Z}_2$ orbifolds, where each $\mathbb{Z}_2$ factor realises a freely-acting K3,  T-duality is broken according to \eqref{Tdbreaking} and the decompactification problem is not present. However, as discussed in \cite{Faraggi:2014eoa}, chirality is then inevitably lost\footnote{A different way to resolve the decompactification problem, which does not spoil chirality and is based on a cancellation between the universal and running parts of the gauge thresholds, was recently considered in \cite{Florakis:2017ecd}.}. This is due to the fact that, although projected with respect to the full $\mathbb{Z}_2 \times \mathbb{Z}_2$ group, the untwisted sector of this orbifold is non chiral, while the three twisted sectors only involve partial $\mathbb{Z}_2$ projectors which yield an enhanced $\mathcal{N}=2$ supersymmetry. Indeed, on the one hand, the vertex operators of the untwisted fermions are in representations of the full ten-dimensional Lorentz group, upon which the $\mathbb{Z}_2$'s have a real action, thus being unable to distinguish between  four-dimensional chiralities. On the other hand, the simultaneous action of twists and shifts is incompatible with a full projection in the twisted sectors, since the independent twisted modular orbit vanishes identically.

The apparent incompatibility between the breaking \eqref{Tdbreaking} and chirality can be lifted if the orbifold group $\Gamma$ has a complex action on the untwisted fermions, so that different spinorial weights carry different charges with respect to $\Gamma$, generating chirality in the untwisted sector itself. To this end, we present a  $T^6 / \mathbb{Z}_3 \times \mathbb{Z}_3^\prime$ orbifold construction where the decompactification problem is absent, the spectrum enjoys four-dimensional chirality, and the string unification scale can be lowered to match the GUT scale.

The first $\mathbb{Z}_3$ factor acts on the three complex coordinates $z_i$ of the factorised $T^6= T^2 \times T^2 \times T^2$, with fixed complex structure $U_i = e^{2  \pi i /6}$, as rotations generated by $v=(\frac{1}{3}, \frac{1}{3} , \frac{2}{3}) $, and corresponds to the celebrated  $Z$ orbifold \cite{Dixon:1985jw} with standard embedding and vanishing Wilson lines. It preserves four supercharges, the gauge group is broken to $\text{E}_6 \times \text{SU}(3) \times \text{E}_8$ and the chiral spectrum comprises 3 copies of chiral multiplets in the representation $(\textbf{27},\textbf{3}, \textbf{1})$ from the untwisted sector as well as 27 copies in $(\textbf{27},\textbf{1}, \textbf{1})$ and 81 copies in $(\textbf{1},\bar{\textbf{3}}, \textbf{1})$. Since $v$ rotates all three $z_i$, the gauge thresholds are $c$-numbers independent of the compactification moduli. 

The second $\mathbb{Z}_3^\prime$ factor, with generator $w=(\frac{1}{3} +\delta , -\frac{1}{3} + \delta , \delta )$, acts as simultaneous opposite rotations of the first two $T^2$'s and order-three shifts $\delta :\ z_i \to z_i + \frac{1}{3} (1+U_i)$ on all three complex coordinates, thus preserving eight supercharges. Notice that the shift $\delta$ is a symmetry of the tori with complex structure $U_i = e^{2  \pi i /6}$ since it permutes the three points $\hat z_i^{(k)} = \frac{k}{3} (1+U_i)$, $k=0,1,2$, of each $T^2$,  fixed under the $2\pi /3$ rotation.

The full orbifold group is thus comprised of the nine elements
\begin{equation}
(1+v+v^2 )(1+w+w^2) = 1+ v+ v^2+ w+ w^2+ v\, w +  v^2\, w^2+ v^2 \, w + v \, w^2 \,,
\end{equation}
out of which one can recognise four independent $\mathbb{Z}_3$ subgroups,  associated to $v$, $w$, $v\, w$ and $v^2\, w$, respectively. While $v$ clearly generates the $T^6/\mathbb{Z}_3$ Calabi-Yau, the others generate three freely-acting K3's which all meet the condition \eqref{Tdbreaking}. Again, the disconnected, twisted, modular orbits are absent in this $T^6/\mathbb{Z}_3 \times \mathbb{Z}_3^\prime$ compactification because $\delta$ permutes the fixed points. Nevertheless, four-dimensional chirality is now preserved both in the untwisted and in the $v$-twisted sectors. 

The spectrum enjoys $\mathcal{N}=1$ supersymmetry, with a non-Abelian gauge group $\text{E}_6 \times \text{E}_8$ and charged matter in the chiral representations $3\times (\textbf{27},\textbf{1})$ from the untwisted sectors as well as in the chiral representations $9\times (\textbf{27}, \textbf{1})$ from the $v$-twisted sector. The twisted sectors associated to the freely-acting orbifolds are instead all massive. 

Gauge thresholds receive moduli-dependent contributions only from the three freely-acting orbifolds and the residual T-duality group  is $\prod_{i=1}^3 \Gamma^1 (3)_{T_i}$, acting on the K\"ahler moduli space. Performing the partial unfolding of \cite{Angelantonj:2013eja}, we may cast the gauge thresholds as
\begin{equation}
\begin{split}
\Delta_{\text{E}_8} &= \sum_{i=1,2,3} \left( \hat Y ^{(i)} -20 \hat \Delta^{(i)} \right) + d_8\,,
\\
\Delta_{\text{E}_6} &= \sum_{i=1,2,3} \left( \hat Y ^{(i)} -8 \hat \Delta^{(i)} \right) + d_{6}\,, 
\end{split}
\end{equation}
where $d_8$ and $d_{6}$ are $c$-number contributions from the $Z$ orbifold, while
\begin{equation}
\begin{split}
\hat Y^{(i)} &= \frac{1}{144} \int_{\mathcal{F}_3} \frac{d^2\tau}{\tau_2^2} \Gamma_{2,2} \big[{\textstyle{0\atop 1}}\big] (T_i , U_i )  \Biggl[ \frac{\hat E_2 E_4 ( 3 E_4 X_3 - 2 E_6 ) }{2\, \eta^{24}}
\\
&\qquad\qquad\qquad\qquad\qquad\qquad + \frac{  E_4 ( 2 E_4^2 - 3 X_3 E_6 ) }{2\, \eta^{24}} +1152 \Biggr]\,,
\end{split}
\end{equation}
and 
\begin{equation}
\hat \Delta^{(i)} = \int_{\mathcal{F}_3} \frac{d^2\tau}{\tau_2^2} \Gamma_{2,2} \big[{\textstyle{0\atop 1}}\big] (T_i , U_i )\,,
\end{equation}
involve the Narain lattice of eq. \eqref{Narainshifted} with the momentum shift  $(\lambda_1 , \lambda_2 ) = (1,1)$.  Aside from the Dedekind function and the Eisenstein series of $\text{SL} (2;\mathbb{Z})$, the  $\Gamma_0 (3)$ modular integrals involve the holomorphic weight-two $\Gamma_0 (3)$ modular form $X_3 = E_2 (\tau ) - 3 E_2 (3 \tau )$. 

The integrals in $\hat \Delta ^{(i)}$'s can be evaluated by standard methods and read
\begin{equation}
\begin{split}
\hat\Delta^{(i)} (T_i , U_i ) &= -\log \left[ \frac{\alpha}{3^3} \, T_{i,2} U_{i,2} \left| \frac{\eta^3 (T_i/3)}{\eta (T_i)}\, \frac{\eta^3 (\frac{1+U_i}{3})}{\eta (U_i)} \right|^2 \right]\,,
\\
&\simeq - \log \, \left(\frac{\alpha}{3^3}\, T_{i,2} \, f_3 (U_i )\right) + O( e^{-2 \pi T_2/3})\,,
\end{split}
\end{equation}
and indeed only exhibits logarithmic growth as advocated in \eqref{Tdbreaking}.

A convenient procedure for the evaluation of the universal parts  $\hat Y ^{(i)}$ amounts to representing the quasi-holomorphic functions in the integrands as linear combinations of Niebur-Poincar\'e series  $\mathcal F_\frak{a}(s,\kappa,0)$
associated to the two cusps $\mathfrak{a}=0,\infty$ of $\Gamma_0 (3)$ \cite{Angelantonj:2013eja},
\begin{equation}
\begin{split}
\frac{E_4 (2 E_4^2-3 X_3 E_6)}{2\, \eta^{24}} &= -36\mathcal F_0^{(3)}(1,1,0)-12 \mathcal F_0^{(3)}(1,2,0)+4\mathcal F_\infty^{(3)}(1,1,0)+720\,,
\\
\frac{\hat{ E}_2  E_4 (3 E_4  X_3-2  E_6)}{2\, \eta^{24}} &= 20\mathcal F_\infty^{(3)}(1,1,0)-4\mathcal F_\infty^{(3)}(2,1,0)-36\mathcal F_0^{(3)}(1,2,0)
\\
& +8\mathcal F_0^{(3)}(2,2,0)-36\mathcal F_0^{(3)}(1,1,0)+12\mathcal F_0^{(3)}(2,1,0)+144 \,.
\end{split}\label{NPdec}
\end{equation}
The fundamental domain $\mathcal{F}_3$ may then be unfolded against the Niebur-Poincar\'e series according to the methods developed in \cite{Angelantonj:2013eja, Angelantonj:2015rxa}, and the above integrals can be then explicitly evaluated. The resulting expression is rather cumbersome and involves combinations of real-analytic modular functions of the $T$ and $U$ moduli. For the sake of simplicity, we shall only display the contributions to the leading, possibly singular, behaviour at large volumes.

Potentially divergent terms originate either from the Niebur-Poincar\'e series  with  $s=1$, or from the constant terms in the linear decomposition \eqref{NPdec}
\begin{equation}
\begin{split}
\hat Y^{(i)}_\text{singular} &\sim \log\left[ \frac{|j(T)-744|^{1/3}}{|j_\infty(T/3)+3|}\, \left|\frac{j_\infty(T/3)+231}{j_\infty(T/3)-12}\right|^{9} \right] \,,
\end{split}
\end{equation}
where $j_\infty$ is the Klein function of $\Gamma_0(3)$ attached to the cusp at $\infty$. Although it is clear that a logarithmic growth in $T_2$ is absent, as expected from the fact that the universal contributions are always IR finite, the additional absence of linear growth is due to a non-trivial cancellation between the Klein functions. This is a direct consequence of our approach \eqref{Tdbreaking}. In conclusion, $\hat Y^{(i)}_\text{singular}$ is actually finite and decays exponentially for large volume. 

The leading contribution to the universal thresholds, comes instead from integrals involving the Niebur-Poincar\'e series with $s=2$ in the decompositions \eqref{NPdec}, and arises from the zero-mode\footnote{Actually, the function $I_3 (s;\mathfrak{a})$ refers to $\Gamma_0 (3)$ modular integrals involving Niebur-Poincar\'e series associated to the cusp $\mathfrak{a}$ and with generis $s$ but with $\kappa=1$. The result for integer $\kappa >1$ can be obtained by the action of the Hecke operator $H_\kappa^{(3)}$ acting on the $U$ modulus \cite{Angelantonj:2015rxa}.} 
\begin{equation}
I^{(0)}_3 (s;\mathfrak{a}) = 2^{2s}\, \sqrt{4\pi}\, \Gamma (s-\tfrac{1}{2})\, T_2^{1-s} \, E_\mathfrak{a} (s, 0; U)
\end{equation}
of the function $I_3 (s;\mathfrak{a})$ introduced in \cite{Angelantonj:2015rxa}. As a result, the universal threshold decays inversely with the volume,
\begin{equation}
\hat Y^{(i)} (T_i , U_i ) \sim \frac{c_3 ( U_i)}{T_{i,2}} + O (e^{-2\pi T_{i,2}/3})\,,
\end{equation}
where the $c_3 ( U_i)$ are order-one $c$-numbers related to the real-analytic Eisenstein series $E_\mathfrak{a} (s, 0; U)$ associated to the two cusps of the fundamental domain of $\Gamma_0 (3)$. 

We would like to stress, that the large volume behaviour
\begin{equation}
\hat \Delta^{(i)} \sim - \log \, \left(\frac{\alpha}{3^3}\, T_{i,2} \, f_3 (U_i )\right) \,, \qquad \hat Y^{(i)} \sim \frac{c_3 (U_i)}{T_{i,2}}\,,
\end{equation}
is not  an accidental  feature of the specific example we have presented here, but it is actually a {\em generic property} of heterotic compactifications which satisfy the requirement \eqref{Tdbreaking}.

Gravitational thresholds for the $R^2$ term actually enjoy a similar property. In any heterotic string compactification they are given by the integral of the $\hat E_2$ dependent term in the integrand of the universal thresholds $\hat Y$, for instance eq. \eqref{NPdec} in the explicit model considered here. A similar analysis as for $\hat Y$ can be carried out explicitly  and it is straightforward to see that they can grow at most logarithmically with the volume, provided condition \eqref{Tdbreaking} is satisfied.

The unification of gauge couplings at the GUT scale is a phenomenologically appealing possibility which has already been addressed in the string literature. However, past treatments either required non-trivial Wilson lines \cite{Nilles:1997vk,Stieberger:1998yi} or faced the decompactification problem which drives the theory into the non-perturbative regime very close to the GUT scale.

In this work, we have shown that gauge threshold corrections can be generated that allow for a precise matching of the string and GUT scales for both $\mathcal N=1$ and even $\mathcal N=2$ vacua, without imposing conditions which are too restrictive and unnatural. In order for this to occur, the compactification has to break the original T-duality group down to products of $\Gamma^1 (N)$ subgroups so that both gauge and gravitational thresholds can at most experience a logarithmic growth at large volume, which guarantees the validity of the perturbative analysis. The breaking \eqref{Tdbreaking} of the T-duality group can be easily achieved via freely acting orbifolds, although it is conceivable that compactifications on orbifolds of non-factorisable torii might induce a similar effect.  We have worked out an explicit example with a chiral spectrum in four dimensions where this paradigm is, for the first time, at work.




\section*{Acknowledgements}

We would like to warmly thank Ignatios Antoniadis, Roberto Contino, Emilian Dudas, John Rizos, Stephan Stieberger and Enrico Trincherini for enlightening discussions.
C.A. would like to thank  the CERN Theory Division and I.F. would like to thank the CERN Theory Division as well as the Physics Department of the University of Torino for their kind hospitality during various stages of this work.

\bibliographystyle{utphys}
\providecommand{\href}[2]{#2}\begingroup\raggedright\endgroup

\end{document}